\documentclass[twocolumn,prb,showpacs]{revtex4}
\usepackage{graphicx}
\usepackage{amssymb}
\usepackage{bm}

\begin{document}

\title{Size Effect of Ruderman-Kittel-Kasuya-Yosida Interaction Mediated by
Electrons in Nanoribbons}
\author{Shuo Mi}
\affiliation{Department of Physics, Jilin University, Changchun 130012, China}
\author{Shuo-Hong Yuan}
\affiliation{Department of Physics, Jilin University, Changchun 130012, China}
\author{Pin Lyu}
\affiliation{Department of Physics, Jilin University, Changchun 130012, China}
\date{29 November 2010}

\begin{abstract}
We calculated the Ruderman-Kittel-Kasuya-Yosida (RKKY) interaction between
the magnetic impurities mediated by electrons in nanoribbons. It was shown
that the RKKY interaction is strongly dependent on the width of the
nanoribbon and the transverse positions of the impurities. The transverse
confinement of electrons is responsible for the above size effect of the
RKKY interaction. It provides a potential way to control the RKKY
interaction by changing nanostructure geometry.

\end{abstract}

\maketitle


\section{introduction}

Recent years there has been a renewed interest in the
Ruderman-Kittel-Kasuya-Yosida (RKKY) magnetic interaction \cite%
{kittel1,kittel3, kittel2,review} due to its important role in giant
magnetoresistance in multilayer structures,\cite{gmr} and
ferromagnetism in diluted magnetic semiconductors.\cite{ohno} More
recently the controllable RKKY interaction attracted much attentions
in the field of spintronics and quantum information
processing.\cite{perspective} The two quantized states of the spin
of single localized electron can be considered as a quantum bit, and
the extended nature of the controllable RKKY interaction between the
coupled local spins has the potential application in building
large-scale spin-based quantum computing and quantum computers.
Craig \textit{et al}.\ experimentally demonstrate the
gate-controllable RKKY interaction between the localized spins in
two quantum dots, each in contact with two-dimensional electron
gas.\cite{craig} Earlier an optical technique to generate and
control the RKKY interaction between charged quantum dots was
proposed by using external laser field.\cite{sham} The other
possibility of gate-controllable RKKY interaction mediated by
electrons in the presence of Rashba spin-orbit coupling or by the
helical edge states in quantum spin Hall systems were also analyzed
intensively.\cite{Ima,Pin,Lai,Chang,Xie} The advantage of the
controllable RKKY interaction in devices has stimulated to fully
investigate and understand its properties both theoretically and
experimentally.

The conventional RKKY magnetic interaction between nuclear spins or between
localized spins in metals is mediated by conduction electrons, where there
is no any confinement on electrons. The controllable RKKY interaction mostly
involves semiconductor nanostructures. In the semiconductor nanostructures,
the electrons are confined at least in one or two dimensions vertically to
the electron movement. In this paper, we focus on the size effect of the
RKKY interaction between two localized spins induced by the transversely
confined electrons in nanoribbons. It provides full understanding of the
RKKY interaction in nanostructures and a potential way to control it by
changing nanostructure geometry is pointed out.

The paper is organized as follows. In Sec.\ II, the RKKY interaction
mediated by electrons confined in nanoribbons is derived. In Sec.\ III, we
present and discuss the size effect of the RKKY interaction. Finally we
conclude with a brief summary in Sec.\ IV.

\section{formalism}

We consider two magnetic impurities with the localized spins $\mathbf{S}_{i}$
($i=1,2$) embedded in semiconductor nanoribbons. The electrons in nanoribbon
are itinerant in $x$ direction and confined in the width $d$ in $y$
direction. The localized spins interact with the conduction electrons via
the $s$-$d$ coupling. The Hamiltonian describing the above basic physics is
written as
\begin{eqnarray}
H &=&\sum_{i=1}^{N}\left[ \frac{\hbar ^{2}\mathbf{k}_{i}^{2}}{2m^{\ast }}
+U(y_{i})\right] I-J\sum_{i=1,2}\mathbf{\sigma }_{i}\cdot \mathbf{S}_{i}
\end{eqnarray}
where the first term of the Hamiltonian describes the conduction electrons
moving along $x$ direction and confined in the $y$ direction in the
nanoribbon. $U(y_{i})$ is assumed to be infinite square well potential. The
second term is the $s$-$d$ interaction between the conduction electrons and
the localized spin $\mathbf{S}_{i}$. $J$ is the $s$-$d$ interaction
strength. $\sigma _{i}$ is Pauli matrices, $m^{*}$ is the effective mass of
the conduction electrons, and $\mathbf{k}_{i}$ is the wave vector along $x$
direction.

The eigenvalue and eigenfunction of the single-particle Hamiltonian are
given by
\[
\varepsilon _{nk\sigma }=\frac{\hbar ^{2}\pi ^{2}n^{2}}{2m^{\ast }d^{2}}+%
\frac{\hbar ^{2}k^{2}}{2m^{\ast }},
\]
and
\begin{eqnarray}
\Psi _{n,\mathbf{k,}\sigma }= \Phi _{n}(y)\Phi _{\mathbf{k}}(x)\eta _{\sigma
},
\end{eqnarray}
with $\Phi _{n}(y)=\sqrt\frac{2}{d}\sin(\frac{n\pi}{d} y),$ $\Phi_{\mathbf{k}%
}(x)=\frac{1}{\sqrt{L}}e^{ikx} $, where $L$ is the length of the
quasi-one-dimensional system, and $\eta _{\sigma}$ is the spinor for
electron spin.

In the second quantization representation, the Hamiltonian $H$ is written as
\begin{eqnarray}
H &=&\sum_{nk\mathbf{\sigma }}\varepsilon _{nk}c_{nk\sigma
}^{\dagger}c_{nk\sigma }-\frac{J}{L}\sum_{i=1,2}\sum_{mnkq\mathbf{\mu \nu }%
}\Phi _{m}^{\ast }(y_{i})\Phi _{n}(y_{i})  \nonumber \\
&&\times e^{-i(k-q)x_{i}}\eta _{\mu }^{+}\mathbf{\sigma }_{i}\eta _{\nu
}\cdot \mathbf{S}_{i}c_{mk\mu }^{\dagger}c_{nq\nu },  \label{h3}
\end{eqnarray}
where $c^{\dagger}_{nk\sigma}$ and $c_{nk\sigma}$ are the creation and
annihilation operators of the conduction electron with wave vector $\mathbf{k%
}$ and subband index ${n}$. ${x}_{i}$ is the position of the $i$th localized
spin in $x$ direction.

Next we derive the RKKY interaction by using the second-order perturbation
theory.\cite{method} The effective Hamiltonian of RKKY interaction is given
by
\begin{equation}
H_{\text{RKKY}}=\sum_{\Gamma }\frac{\left\langle \Gamma _{0}\right\vert
H_{s-d}\left\vert \Gamma \right\rangle \left\langle \Gamma \right\vert
H_{s-d}\left\vert \Gamma _{0}\right\rangle }{E_{\Gamma _{0}}-E_{\Gamma }},
\label{h4}
\end{equation}
where $H_{s-d}$ presents the $s-d$ coupling, $\left\vert \Gamma \right\rangle
$ is the excited state with the energy $E_{\Gamma}$, and $\left\vert \Gamma
_{0} \right\rangle$ is the ground state with the energy $E_{\Gamma_0}$. As
usual, the excited state is taken as one particle-hole excited state $%
\left\vert \Gamma \right\rangle =c_{{n}^{\prime } {q}^{\prime }\mu
}^{\dagger} c_{n q\nu}\left\vert \Gamma _{0}\right\rangle $, where ${n}%
^{\prime}{q}^{\prime}\mu$ and $nq\nu$ satisfy $E_{{n}^{\prime}{q}%
^{\prime}\mu}>E_{F}$ and $E_{n q\nu}<E_{F}$, respectively. From Eqs.\ (\ref%
{h3})-(\ref{h4}), we obtain the effective Hamiltonian
\begin{eqnarray}
H_{\text{RKKY}} &=&J_{\text{eff}}(r,y_1,y_2)\mathbf{S}_{1}\cdot \mathbf{S}%
_{2} \\
J_{\text{eff}}(r,y_1,y_2)&=&\frac{4J^{2}}{L^{2}}\sum_{mn kq} \Phi
_{n}(y_{1})\Phi _{m}(y_{1}) \Phi _{m}(y_{2})\Phi _{n}(y_{2})  \nonumber \\
&&\times \cos [(k-q)r]\frac{f(E_{mk})[1-f(E_{nq})]}{E_{m\mathbf{k}}-E_{nq}}
\end{eqnarray}
where $\mathbf{r}$ is the relative distance vector between the two
localized spins given by where $r=x_{1}-x_{2}$, $f(E_{nk})$ is the
Fermi-Dirac distribution function.

Eq.(6) is the main result of our present paper. The coupling strength $J_{%
\mathrm{eff}}(r,y_1,y_2)$ is dependent on the transverse positions
of the two local spins. It is due to the broken symmetry of
lattice translation in transverse direction. In the longitudinal
direction the translational symmetry hold so that the
$J_{\mathrm{eff}}(r,y_1,y_2)$ depends only on the relative
distance $r$ between the two local spins in $x$ direction. There
is no magnetic coupling between the two local spins when the
transverse positions satisfy $\sin(\frac{n\pi}{d}y)=0$. Another
feature is the different subbands occupied by electrons have
different contributions to the magnetic coupling. This leads to
the summation of different oscillations existing.

The above formula may be generalized to the quasi-two-dimensional system and
it has similar property of the size effect for the RKKY interaction.



\section{result and discussion}

Our numerical results of the RKKY interaction in nanoribbons at zero
temperature is presented below. We chose the following typical
material parameters: \mbox{$m^{*}=0.067m_{e} $}, the density $n_e$
between $1.0\times 10^{11}~\mathrm{cm}^{-2}$ and $3.0\times
10^{11}~\mathrm{cm}^{-2}$ appropriate for two dimensional electron
gas in the semiconductor
heterostructure. Also we used $J=1.0~\mathrm{eV~nm^{2} }$ for the $s$-$%
d$ coupling strength. In all the figures we take $k_{F}r$ as the
dimensionless distance between the two localized spins along $x$
direction, where $k_{F}$ is the Fermi wave vector for the
two-dimensional electron gas.

\begin{figure}[tbp]
\includegraphics[width=8.0cm]{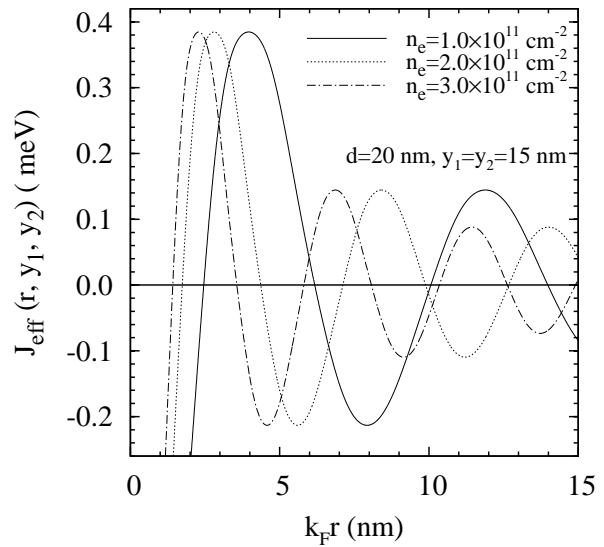}
\caption{Indirect RKKY magnetic interaction strength versus the
distance between two localized spins along $x$ direction at several
different electron densities with fixed width \mbox{$d=20$ nm}.}
\label{fig1}
\end{figure}

\begin{figure}[tbp]
\includegraphics[width=8.0cm]{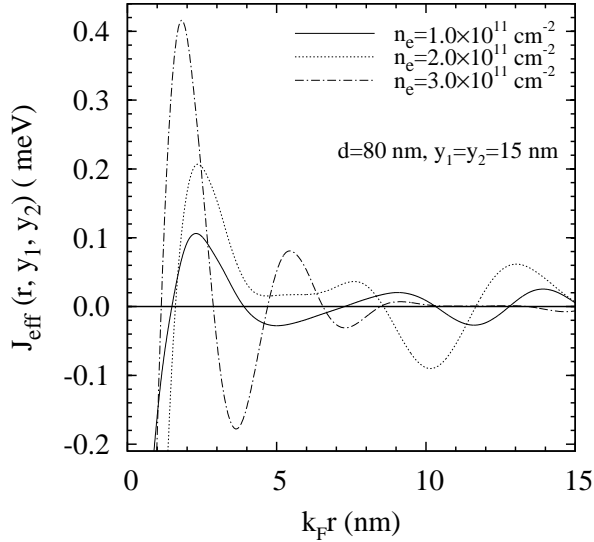}
\caption{Indirect RKKY magnetic interaction strength versus the
distance between two localized spins along $x$ direction at several
different electron densities with fixed width \mbox{$d=80$ nm}.}
\label{fig2}
\end{figure}

The RKKY interaction energy $J_{\mathrm{eff}}$ versus $k_{F}r$ at
different electron densities for the fixed nanoribbon width $d=20$
nm and $d=80$ nm are plotted in Fig.\ \ref{fig1} and \mbox{Fig.\
\ref{fig2}}, respectively. The behavior of the RKKY interaction
strength for $d=20$ nm is similar to that of the conventional RKKY
interaction such as oscillation and decay. This is because there is
only one occupied subband for the present $d=20$ nm confined system.
The main difference between the conventional and present situations
is that the RKKY interaction strength in nanoribbons is closely
related to the transverse confinement in this nanostructure. When
the width of the system becomes large, there are more subbands
filled by electrons under the Fermi energy level. Each of these
subbands contributes to the RKKY interaction strengths with
different oscillation periods, which lead to the suppression or the
enhancement of the strength after summation, as shown in Fig.\
\ref{fig2} where $d=80$ nm. See the dot line in the area of
\mbox{$k_{F}r=5$ nm}, the platform structure is an evident
suppression of the oscillation. Similar phenomena of the suppression
and the enhancement was also found in the RKKY interaction mediated
by the spin-orbit coupling electrons systems.\cite{Pin} Moreover, it
is obvious that the RKKY interaction strength is related to the
electron density. With a fixed width of the nanoribbon, the larger
the electron density is, the more subbands there are under the Fermi
energy level. Thus the summation of the RKKY interaction strengths
may be more complicated. For example in Fig.\ \ref{fig2} the chain
line in the vicinity of $k_{F}r=11$ nm shows that the RKKY
interaction strength is approaching to zero while in the vicinity of
\mbox{$k_{F}r= 15$ nm} the strength turns out to have an obvious
enhanced structure. Such cases of firstly-damping-then-enhancing
structures can by no means take place in the conventional RKKY
interaction. We ascribe this phenomenon to the coupling of subbands
interactions with different periods. At last, we should point out
that it is not necessary as shown in  Fig.\ \ref{fig2} that the
strength amplitude is increasing with the electron density. We found
that in certain situations, the interaction of subbands could weaken
RKKY interaction strength in our model.

\begin{figure}[tbp]
\includegraphics[width=8.0cm]{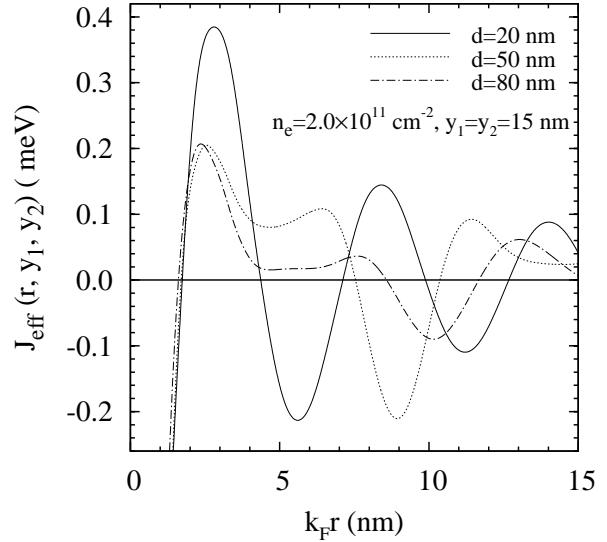}
\caption{Indirect RKKY magnetic interaction strength versus the distance
between two localized spins at several different width of nanoribbons under
the same electron density \mbox{$2.0\times 10^{11}
\rm{cm^{-2}}$.}}
\label{fig3}
\end{figure}

The dependence of the RKKY interaction on the width $d$ is more
clearly demonstrated in Fig.\ \ref{fig3}. In this case, we fixed the
electron density $n_e$, and the positions of the two impurities
$y_1$ and $y_2$. Corresponding to the different widths $d=20$ nm,
$50$ nm, and $80$ nm, there are one, two, and three occupied
subbands under Fermi energy, respectively. For the situation $d=20$
nm, the
numerical results goes back to the conventional RKKY interactions. As to $%
d=50$ nm, the final RKKY interaction strength is the summation of
four terms concerning both intro-subbands and inter-subbands
interactions with each term represents an independent oscillation
mode and decaying tendency. A more complicated situation of the
RKKY interaction originated from nine term contributions takes
place when $d=80$ nm. No matter how complex the situation is, the
trend of the decaying oscillation still holds although the RKKY
interaction strength may not have regular period.


\begin{figure}[tbp]
\includegraphics[width=8.0cm]{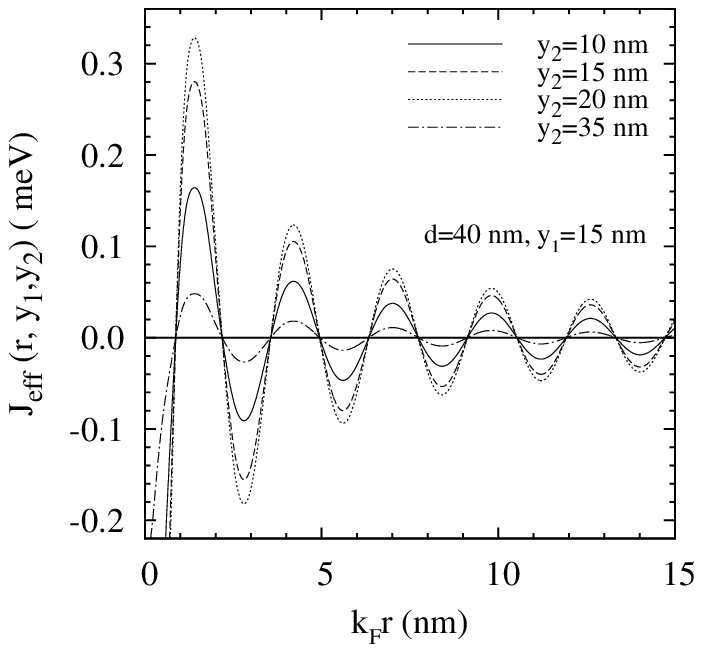}
\caption{Indirect RKKY magnetic interaction strength versus the
distance between two localized spins with nanoriboon width
\mbox{$d=40$ nm}, electron density $n_{e}=2.0 \times 10^{11}
\mathrm{{cm^{-2}}}$. We fixed one impurity at $y_{1}=15$ nm, and
change the location of another impurity $y_{2}$.} \label{fig4}
\end{figure}

\begin{figure}[tbp]
\includegraphics[width=8.0cm]{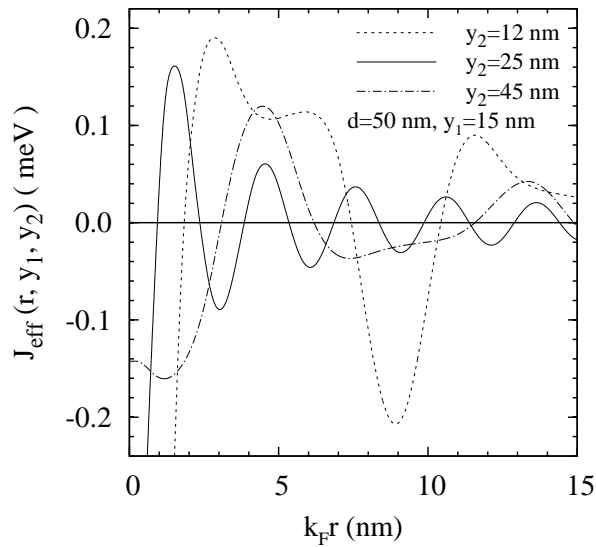}
\caption{Indirect RKKY magnetic interaction strength versus the
distance between two localized spins with nanoribbon width $d=50$
nm, electron density $n_{e}=2.0 \times 10^{11} \mathrm{{cm^{-2}}}$.
We fixed one impurity at $y_{1}=15$ nm, and change the location of
another impurity $y_{2} $.} \label{fig5}
\end{figure}

Another important feature of the RKKY interaction in nanoribbons is
strongly dependent on the transverse position of the impurities as
shown in \mbox{Fig.\ \ref{fig4}} and \mbox{Fig.\ \ref{fig5}}. We fix
the electron density to be $2.0\times10^{11}~\mathrm{{cm^{-2}}}$,
and set the position of one impurity at $y_{1}=15$ nm. By varying
the location $y_{2}$ of the other impurity we conclude as follow. In
Fig.\ \ref{fig4} where $d=40$ nm, only one subband is occupied under
the Fermi Level. Like previous situation, it coincides with the
conventional RKKY interaction, which has strictly periodical
conformity, i.e. the location $y_2$ of the other impurity only
affects the amplitude of the RKKY interaction strength, but has no
influence on the periods of the oscillation. Moreover, the system
possesses a transverse symmetry for $y_{2}$ due to our using the
infinite quantum well as the transverse confinement on the
nanoribbon in model building. This means for example that $y_{2}=10$
nm and $y_{2}=30$ nm share an identical figure. Same to
\mbox{$y_{2}=15$ nm} and \mbox{$y_{2}=25$ nm}. However, in other
situations when there are more subbands under the Fermi Level for
instance $d=50$ nm, see Fig.\ \ref{fig5}, the existence of subbands
interactions bring about the transverse symmetry breaking of $y_2$.
This can also be verified by one of the curves, see the dash line
corresponding to $y_{2}=25$ nm in Fig.\ \ref{fig5}, which settles in
the very middle of the nanoribbon. The figure is special in that it
conforms with the conventional RKKY interaction strength. This
result can be attributed to the special location of $y_{2}$ which
leads to only one subband interaction term to be nonzero in the
summation. This subband contribution alone is a conventional RKKY
interaction form. Similar results are also spotted in other
nanoribbon systems with different width.

\section{summary}

We derived the RKKY interaction mediated by electrons in nanoribbon. Our
theoretical results demonstrate that the RKKY interaction is strongly
dependent on the width of the nanoribbon and the positions of impurities in
transverse direction. The transverse confinement of electrons is responsible
to the size effect of the RKKY interaction. It provides the potential way to
control the RKKY interaction between the local spins by tuning geometry.

\section{acknowledgment}

This work was supported by the National Science Foundation of China grant
No.60976072.

\end{document}